\documentclass[conference]{IEEEtran}
\usepackage{graphicx}
\usepackage{hyperref}
\usepackage{amsmath}
\usepackage{listings}
\usepackage{comment}
\usepackage[backend=biber]{biblatex}
\usepackage[font=bf]{caption}

\begin{document}

\title{MARCO: Multi-Agent Code Optimization with Real-Time Knowledge Integration for High-Performance Computing}

\author{
    \IEEEauthorblockN{Asif Rahman, Veljko Cvetkovic, Kathleen Reece, Aidan Walters,\\ Yasir Hassan, Aneesh Tummeti, Bryan Torres, Denise Cooney, Margaret Ellis, and Dimitrios S. Nikolopoulos}
    \IEEEauthorblockA{Department of Computer Science, Virginia Tech} 
    \\
    Emails: \{asifr, veljko, katkreece501, aidanw, yasirh, artummeti, bryantorres, denise05, maellis1, dsn\}@vt.edu}

\maketitle

\begin{abstract}
Large language models (LLMs) have transformed software development through code generation capabilities, yet their effectiveness for high-performance computing (HPC) remains limited. HPC code requires specialized optimizations for parallelism, memory efficiency, and architecture-specific considerations that general-purpose LLMs often overlook. We present MARCO (Multi-Agent Reactive Code Optimizer), a novel framework that enhances LLM-generated code for HPC through a specialized multi-agent architecture. MARCO employs separate agents for code generation and performance evaluation, connected by a feedback loop that progressively refines optimizations. A key innovation is MARCO's web-search component that retrieves real-time optimization techniques from recent conference proceedings and research publications, bridging the knowledge gap in pre-trained LLMs. Our extensive evaluation on the LeetCode 75 problem set demonstrates that MARCO achieves a 14.6\% average runtime reduction compared to Claude 3.5 Sonnet alone, while the integration of the web-search component yields a 30.9\% performance improvement over the base MARCO system. These results highlight the potential of multi-agent systems to address the specialized requirements of high-performance code generation, offering a cost-effective alternative to domain-specific model fine-tuning.
\end{abstract}

\section{Introduction}

High-performance computing (HPC) represents the pinnacle of computational power, utilizing clusters of computing resources to overcome the limitations of individual machines. HPC's fundamental advantage lies in implementing parallel processing techniques that maximize processor cluster performance, enabling complex data processing and mathematical calculations that would otherwise be infeasible~\cite{Nichols_2024}. HPC has been instrumental in driving innovation across diverse domains including climate modeling, astrophysics simulations, pharmaceutical research, energy optimization, financial risk analysis, and training state-of-the-art machine learning models, particularly Large Language Models~\cite{chen2024landscapechallengeshpcresearch,ding2023hpc,hoffmann2022training,jin2023evaluation,narayanan2021efficient,yin2023ion}.

Large Language Models (LLMs) have revolutionized technology across multiple domains, democratizing access to sophisticated programming through natural language interfaces. These models have evolved from conversational assistants to autonomous systems capable of executing traditionally manual tasks, such as code generation~\cite{jiang2024survey,wang2024llm,liu2024evaluating,phuong2024TDD,li2024using}. The ability of LLMs to generate, test, and evaluate code has significantly enhanced software engineering capabilities, reducing development cycles and increasing productivity.

However, despite their general utility, LLMs face significant challenges when applied to domain-specific knowledge requirements. According to the ParEval benchmark, both open-source and commercial LLMs demonstrate substantially lower accuracy in generating correct and efficient parallel code compared to serial code~\cite{Nichols_2024}. This limitation stems from LLMs' training methodology, which typically involves diverse internet-sourced data rather than specialized domain knowledge. The relative scarcity of HPC code examples available online further exacerbates this knowledge gap~\cite{Nichols_2024,chen2024landscapechallengeshpcresearch}. While LLMs have also been successfully deployed in parallel program verification~\cite{aaron_iwomp24,sollenberger2024llm4vvexploringllmasajudgevalidation}, the unique demands of high-performance computing—requiring rigorously tested and optimized code for parallelism, memory efficiency, and execution speed—raise a critical research question: How can we effectively leverage LLMs to address HPC-specific programming challenges comprehensively?

A promising approach to addressing LLM limitations is the implementation of Artificial Intelligence (AI) Agents—autonomous LLM-based systems capable of performing complex tasks through customized workflows and specialized tools~\cite{xi2023survey,acharya2024agentic,li2024survey,gupta2024survey,sudalaimuthu2024enhancing}. AI agents extend beyond natural language processing to encompass decision-making, problem-solving, and execution capabilities. By utilizing external tools, these agents can access up-to-date information and effectively complete domain-specific tasks. Multi-agent systems further enhance these capabilities by distributing complex tasks across specialized agents, enabling the handling of work that exceeds the capabilities of any single agent.

\subsection{Research Contributions}

This paper presents MARCO (Multi-Agent Reactive Code Optimizer), a novel framework that fundamentally reimagines code optimization for high-performance computing. Unlike traditional approaches that rely on monolithic LLMs or extensive fine-tuning, MARCO implements a specialized multi-agent architecture with several key innovations:

\begin{itemize}
    \item \textbf{Specialized Agent Architecture:} MARCO introduces a clear separation between code generation, performance evaluation, and testing agents, allowing each to develop distinct expertise. The code generation agent specializes in implementing advanced optimization techniques including cache locality optimization, parallelization through OpenMP or CUDA, and vectorization strategies. Meanwhile, the evaluation agent focuses on rigorously assessing execution time, memory usage, and algorithmic complexity, while the testing agent continuously evaluates output correctness.
    
    \item \textbf{Adaptive Feedback Mechanism:} Unlike traditional single-pass LLM code generation, MARCO implements an iterative feedback loop between its specialized agents. The evaluation agent collects comprehensive performance metrics that inform subsequent optimization attempts, creating a progressive refinement process that consistently improves code quality beyond what conventional approaches can achieve.
    
    \item \textbf{Real-time Knowledge Integration:} MARCO incorporates a novel web-search component that actively retrieves the latest optimization techniques from research publications and conference proceedings, effectively bridging the knowledge gap between pre-trained LLM cutoff dates and current state-of-the-art HPC practices. This dynamic knowledge acquisition represents a significant advancement over static fine-tuning approaches.
    
    \item \textbf{Cost-Effective Implementation:} By strategically separating optimization and evaluation processes, MARCO minimizes token usage and API call volume, offering a more economical alternative to expensive model fine-tuning or training specialized HPC models from scratch.
\end{itemize}

To assess the effectiveness of our approach, we first assembled a comprehensive evaluation dataset, which included the LeetCode 75 problem set along with an additional 10 challenging problems. Next, we implemented MARCO and compared its performance to leading language models such as GPT-4o, Claude 3.5 Sonnet, DeepSeek Coder-V2, and Llama 3.1. Additionally, we conducted detailed ablation studies to measure the contribution of each component within the MARCO architecture.

Our extensive experimental evaluation demonstrates that MARCO achieves a significant 14.6\% reduction in average runtime on the LeetCode 75 problem set compared to Claude 3.5 Sonnet alone, while the integration of the web-search component yields a 30.9\% performance improvement over the base MARCO system. These results establish MARCO as a transformative approach to HPC code optimization that outperforms current state-of-the-art models without requiring specialized training or fine-tuning.

The remainder of this paper is organized as follows: Section~\ref{background} discusses background and related work, Section~\ref{methodology} details our methodology, Section~\ref{results} presents experimental results, Section~\ref{marco} provides an in-depth analysis of the MARCO system, Section~\ref{discussion} discusses our findings, and Section~\ref{conclusions} concludes with future research directions.
\section{Background \& Related Work}
\label{background}

\subsection{LLMs for Code Generation}

LLMs can perform a wide range of tasks, such as code generation and understanding, through their parameters and transformer architecture~\cite{wang2024llm,liu2024evaluating,phuong2024TDD,li2024survey}. LLMs contain billions of parameters that enable them to identify patterns in the data encountered during training. Research demonstrates a strong correlation between parameter size and model effectiveness; as the number of parameters increases, the model's ability to solve higher-complexity or data-intensive problems generally improves~\cite{kaplan2020scaling,hoffmann2022training,earle2023level}. The transformer architecture, a specialized type of neural network, is designed to process sequential data and predict intended outcomes through attention mechanisms.

Chatbots like ChatGPT were among the first to provide general-purpose assistance across numerous domains. LLMs acquire domain knowledge through training on large datasets spanning diverse fields. OpenAI, for example, trained ChatGPT using techniques like Reinforcement Learning from Human Feedback~\cite{ouyang2022training,christiano2017deep,liu2024best}, which utilizes a reward system to reinforce positive responses based on human trainer feedback. Given the adaptability of LLMs, specialization has emerged through the development of domain-specific models designed for particular applications. Code generation represents one popular specialization domain, where models are pre-trained primarily on code repositories to enhance their programming capabilities. Advanced code-focused LLMs like Qwen2.5-Coder employ techniques such as synthetic data generation to address the scarcity of high-quality code examples~\cite{hui2024qwen,wang2024llm}.

While LLMs have demonstrated impressive performance in code generation tasks, their effectiveness is primarily limited to small-scale problems, such as individual functions and single-file programs~\cite{li2024performance}. As coding problems increase in complexity and scale, these models experience significant performance degradation, largely due to limitations in their context processing capabilities~\cite{wang2024large}. This degradation relates directly to the LLM's context window—the maximum amount of information the model can consider when generating responses, measured in tokens. LLMs process natural language by segmenting it into tokens, which can range from individual characters to complete words, enabling them to interpret and generate human-like text~\cite{balfroid2024towards}. Research has demonstrated that context window size critically determines an LLM's effectiveness for complex tasks like code generation, as larger windows enable models to process more documentation and maintain awareness of broader code structures~\cite{antropic2023introducing}. This capability has evolved significantly, progressing from earlier models like GPT-3.5 with a context window of 16,385 tokens to more advanced models such as GPT-4 Turbo, which supports substantially expanded context windows~\cite{li2024performance}.

\subsection{Challenges in LLM-Generated HPC Code}

LLM-generated code often exhibits significant deficiencies in performance optimization, including inefficient cache utilization and suboptimal parallelization strategies~\cite{nichols2024large}. The ParEval benchmark has revealed a substantial gap between LLMs' ability to generate serial versus parallel code, with all evaluated models demonstrating markedly lower accuracy when generating parallel implementations~\cite{nichols2024large, davis2024pareval}. While open-source repositories provide vast amounts of training data for general coding tasks, this breadth comes at the expense of the specialized domain knowledge required for high-performance computing applications~\cite{ding2023hpc}. The primary challenge stems from how LLMs process information through Natural Language Processing (NLP) techniques, which excel with textual inputs but struggle with the computationally intensive scientific data prevalent in HPC contexts~\cite{chen2023lm4hpc}.

Modern HPC infrastructure leverages parallel execution on multi-core systems through threading and synchronization mechanisms to optimize performance. However, due to the relatively recent emergence of these architectures, LLMs have limited understanding of modern parallel programming paradigms~\cite{kadosh2023scope}. This knowledge gap results in parallel code generation that performs substantially worse than sequential implementations~\cite{nichols2024large}. Furthermore, the advanced algorithms and optimization techniques typical in HPC environments create additional complexity that makes error diagnosis and correction particularly challenging for current LLM implementations~\cite{ding2023hpc}.

These contextually demanding factors cause general-purpose LLMs to struggle with comprehensive HPC optimization. Recent research has begun addressing this problem with the introduction of HPC-Coder, which fine-tunes pre-existing models on HPC-specific data to improve parallel code generation capabilities~\cite{nichols2023hpc}. This specialized model outperforms general-purpose LLMs on HPC-specific tasks and OpenMP pragma generation, demonstrating the potential for targeted HPC optimization in LLM applications.

LLM-generated code also struggles with complex problems on initial attempts, often requiring multiple iterations to achieve correctness~\cite{huang2023agentcoder}. LLMs are inherently limited by single-pass generation approaches, which may be insufficient for challenging programming tasks. An agent-based approach offers a promising alternative for complex code generation, allowing multiple attempts at problem-solving and iterative code improvement~\cite{dong2023selfcollab}. This mirrors human problem-solving behavior, where programmers identify errors, learn from mistakes, and refine their solutions~\cite{yuqian2024colearning}. Given the limitations of LLMs in HPC code generation, simple single-model approaches may be inadequate for addressing these specialized requirements~\cite{singh2023selfplanning}.
\section{Methodology}
\label{methodology}

\subsection{Problem Statement and Approach}

Current approaches to code generation using Large Language Models (LLMs) exhibit significant limitations when applied to high-performance computing (HPC) workloads. Most existing models generate code in a single-pass manner, focusing primarily on syntactic correctness rather than optimizing for parallelism, memory hierarchy, or target machine characteristics. As a result, generated code often underperforms in memory efficiency, execution speed, and scalability, especially compared to hand-optimized or domain-specific implementations. This performance gap requires substantial manual tuning and expert intervention, highlighting the inadequacy of current LLM-driven workflows for HPC applications. Recent studies, including HPC-GPT~\cite{ding2023hpc} and the ParEval benchmark~\cite{nichols2024large}, have empirically demonstrated that LLMs struggle to deliver performant parallel code, revealing a clear need for more sophisticated, architecture-aware generation techniques. Furthermore, work by Nichols et al.~\cite{nichols2023hpc} shows that even fine-tuned models on HPC-specific data require additional optimization layers to approach expert-written code performance.

To address these challenges, we introduce \textbf{MARCO} (Multi-Agent Reactive Code Optimizer), a multi-agent system that fundamentally reimagines the code optimization process by combining automated code generation with systematic performance testing through API integrations for dynamic adaptation. This modular separation between generation and evaluation enables improvements beyond conventional monolithic approaches. MARCO's architecture centers on four key innovations:

\begin{itemize}
    \item \textbf{Specialized Agent Separation:} MARCO implements a clear division of responsibilities between the code optimizer and testing agent. The optimizer focuses on applying advanced strategies such as cache locality improvements, parallelization through OpenMP or CUDA, and vectorization techniques, drawing upon insights from recent high-performance computing literature~\cite{ding2023hpc,nichols2023hpc}. Meanwhile, the testing agent rigorously evaluates execution time, memory consumption, algorithmic complexity, output correctness, and incremental improvements over the original unoptimized code.
    
    \item \textbf{Adaptive Feedback Loop:} Unlike traditional single-pass LLM systems that seek immediate solutions, MARCO employs an iterative feedback mechanism. Performance metrics gathered by the testing agent are fed back to the optimizer, creating a refinement loop that progressively drives the system toward more efficient implementations, aligning with multi-step agentic workflows recently explored in AI systems research~\cite{huang2023agentcoder}.
    
    \item \textbf{Cost-Efficient Design:} By decoupling optimization and evaluation tasks, MARCO reduces unnecessary API calls and minimizes token consumption at each iteration. This architectural design ensures controlled system costs, offering a scalable, resource-aware alternative to computationally expensive fine-tuning approaches.
    
    \item \textbf{Broad Applicability:} While many existing frameworks rely on specialized fine-tuning for narrow HPC tasks, MARCO dynamically adapts to live performance data and integrates real-time insights from web-sourced research. This generalizable design reduces the time and resources required for model retraining, maintaining efficiency across diverse applications.
\end{itemize}

\subsection{Selected LLMs}

We evaluated a range of leading proprietary and open-source LLMs for their capacity to generate high-performance computing code, including GPT-4o, Claude 3.5 Sonnet, DeepSeek Coder-V2, and Llama 3.1 8b. GPT-4o~\cite{openai_gpt4o} and Claude~\cite{anthropic_claude} are proprietary models renowned for their robust infrastructure, offering high-throughput performance and accessible APIs for integration into external systems. In contrast, DeepSeek Coder-V2~\cite{deepseek_coder} and Llama 3.1 8b~\cite{llama3_2024} are prominent open-source models that enable local deployment, extensive customization, and fine-grained control over parameters, memory allocation, and inference settings.

For MARCO's implementation, we selected the ChatGPT (OpenAI) and Claude (Anthropic) APIs as primary LLM providers due to their mature developer ecosystems, comprehensive performance telemetry available through developer consoles, and seamless integration with auxiliary technologies required in multi-agent workflows.

\subsection{Benchmark Design}

To rigorously assess MARCO's optimization capabilities, we designed a benchmark suite comprising tasks that reflect core challenges in high-performance computing. These tasks span three primary categories:

\begin{itemize}
    \item \textbf{Parallelization:} We tested MARCO's ability to generate and optimize code using OpenMP, CUDA, MPI, and Python's \texttt{asyncio}, enabling concurrency within single-threaded contexts and across multi-threaded or multi-GPU environments. Our evaluation framework carefully manages thread creation to avoid performance degradation in scenarios with small workloads, a common pitfall in parallel code generation~\cite{ding2023hpc,nichols2024large}.
    
    \item \textbf{Memory Optimization:} We evaluated MARCO's implementation of vectorization and cache utilization strategies to improve data locality and reduce memory access bottlenecks. The system leverages its integrated web-search component to retrieve cutting-edge optimization techniques from research publications indexed in IEEE, ACM, and arXiv, addressing the knowledge gap in pretrained LLMs~\cite{nichols2023hpc}.
    
    \item \textbf{Algorithmic Efficiency:} We examined classical algorithmic problems, including sorting and matrix multiplication, using MARCO's testing agent to provide detailed runtime and space complexity comparisons for each improvement over baseline implementations. This enables users to make informed, data-driven decisions about optimization strategy adoption.
\end{itemize}

\subsection{Evaluation Metrics}

To comprehensively assess MARCO's performance and utility, we adopted a multi-dimensional evaluation framework spanning execution performance, code quality, and cost-efficiency:

\begin{itemize}
    \item \textbf{Performance Metrics:} We measured three key performance indicators:
    \begin{itemize}
        \item \textit{Execution time} was measured using system-level profiling tools, such as \texttt{time.h} in C and high-resolution timers in Python, following best practices established in LLM benchmarking literature~\cite{li2024performance}.
        \item \textit{Computational throughput} was assessed using FLOPS measurements to enable direct comparisons across different optimization strategies.
        \item \textit{Memory usage} was evaluated using Python packages including \texttt{tracemalloc} for tracking high-memory objects and \texttt{psutil} for monitoring system-level resource utilization, consistent with techniques reported in recent HPC optimization studies~\cite{ding2023hpc,yin2023ion}.
    \end{itemize}
    
    \item \textbf{Code Quality Assessment:} We evaluated two primary aspects:
    \begin{itemize}
        \item \textit{Correctness} was validated by submitting generated solutions to the LeetCode problem framework, which provides automated feedback on test case performance. While certain limitations existed due to library import restrictions, this approach aligned with established methods in prior code-generation evaluations~\cite{li2024performance}.
        \item \textit{Readability and maintainability} were preserved through MARCO's approach of making localized improvements rather than complete rewrites. All modifications were accompanied by inline comments and high-level explanations to maintain developer transparency.
    \end{itemize}

    \item \textbf{Cost Analysis:} We measured API cost per query and token consumption. During March and April, the Anthropic (Claude) models processed approximately 562,703 input tokens and 213,227 output tokens, incurring a cost of \$4.22 as recorded in the Anthropic developer console. For OpenAI (ChatGPT) models over February to April, 1,145,708 input tokens and 898,465 output tokens were consumed, at a total cost of \$10.50, yielding an average API cost per query of approximately \$0.00369. These cost assessments are critical for evaluating the practicality of multi-agent frameworks like MARCO, especially relative to large-scale fine-tuning or dedicated model retraining~\cite{nichols2023hpc}.
\end{itemize}
\section{Experimental Results}
\label{results}

\subsection{Experimental Setup}

To ensure reproducibility and provide a fair evaluation of MARCO's performance, all experiments were conducted under controlled hardware and software conditions. Specifically, we ran benchmarks on a machine equipped with an AMD EPYC 7502 32-Core Processor, NVIDIA A100 GPUs, and 256 GB of system memory, operating under Ubuntu 22.04 LTS. The software environment included Python 3.10, CUDA 12.2 for GPU-accelerated tasks, and relevant dependencies including Python's built-in \texttt{tracemalloc} module and \texttt{psutil} version 5.9.8.

Each experiment was repeated 5 times to account for random variability, and reported metrics (execution time, memory usage) reflect the average over these runs, with standard deviations provided where applicable. Random seeds were set to 42 for all stochastic operations to maintain consistency. All code was executed within a \texttt{venv} virtual environment to ensure consistent dependency resolution across experiments. Additionally, Docker containers (python:3.10-slim with CUDA 12.2 support) were used in select experiments to support dynamic library installation and compatibility, particularly when testing code requiring packages beyond the default system installation.

\subsection{Performance Analysis}

\begin{figure}[h]
    \centering
    \includegraphics[width=0.4\textwidth]{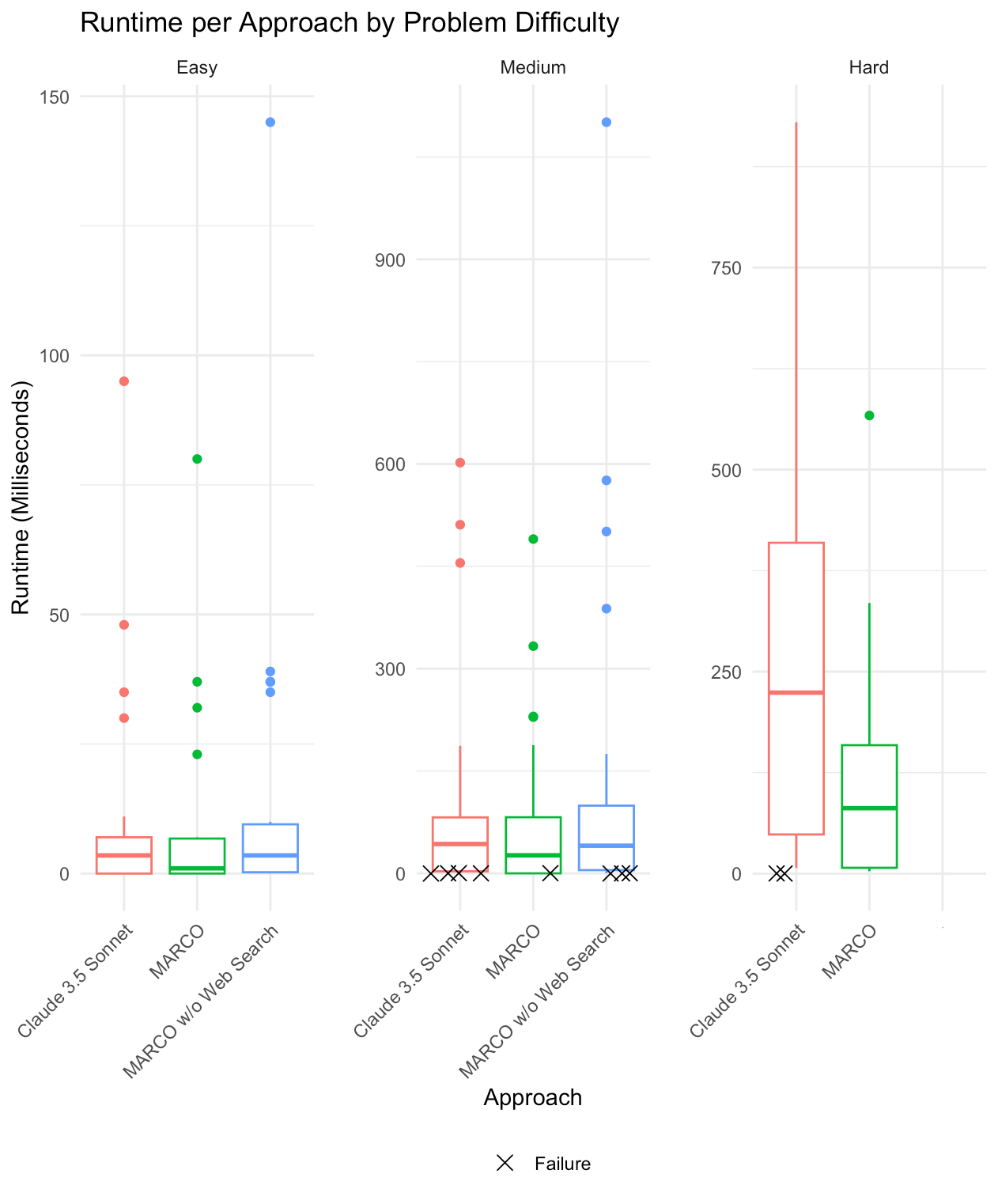}
    \caption{\textbf{Runtime distribution comparison between Easy/Medium difficulty problems from the LeetCode 75 dataset and 10 selected Hard difficulty problems, showing MARCO's performance improvements across difficulty levels.}}
    \label{fig:example}
\end{figure}

We systematically evaluated MARCO-optimized code against baseline LLM-generated code, including Claude-3.5 Sonnet, GPT-4o, DeepSeek Coder-V2, and Llama 3.1 8b, on a diverse set of programming tasks from the LeetCode 75 problem set and 10 additional randomly selected hard problems. Our analysis focused on runtime performance, consistency, and relative improvements. We report detailed results from comparisons with Claude-3.5 Sonnet, which consistently exhibited the highest code performance among the tested LLMs.

On the LeetCode 75 (easy and medium) problems, MARCO matched or outperformed Claude-3.5 Sonnet in 44 out of 75 cases, achieving an average runtime reduction of 14.6\% (mean runtime 49.1 ms vs. 57.5 ms). For hard problems, MARCO demonstrated stronger gains, outperforming Claude-3.5 Sonnet in all 10 cases, with an average runtime improvement of 51.9\%. Notably, MARCO's solutions ranked in the top 20\% of all LeetCode submissions for 7 out of 10 hard problems, demonstrating competitive performance against the broader programming community.

Table~\ref{tab:runtime-results} summarizes average runtime performance across problem difficulties for the two best-performing approaches.

\begin{table}[h]
    \centering
    \caption{Average Runtime (ms) Across Problem Difficulties}
    \label{tab:runtime-results}
    \begin{tabular}{lccc}
    \hline
    \textbf{Model} & \textbf{Easy} & \textbf{Medium} & \textbf{Hard} \\
    \hline
    Claude-3.5 Sonnet & 11.8 & 78.1 & 288.1 \\
    MARCO (ours)      & 9.3 & 65.9 & 138.5 \\
    \textbf{Improvement} & \textbf{21.2\%} & \textbf{15.6\%} & \textbf{51.9\%} \\
    \hline
    \end{tabular}
\end{table}

The results demonstrate that MARCO's multi-agent approach becomes increasingly effective as problem complexity increases, with the most significant improvements observed on hard problems where sophisticated optimization strategies provide the greatest benefit.

\subsection{Memory Usage Analysis}

We compared memory usage between MARCO-optimized and baseline LLM-generated code using peak memory consumption metrics recorded with Python's \texttt{tracemalloc} module and \texttt{psutil}. Among the 85 evaluated problems (75 easy/medium, 10 hard), MARCO showed measurable memory usage improvements in 15 cases, while baseline LLMs showed advantages in 10 cases, with the remaining 60 cases showing comparable performance within measurement variance.

On average, MARCO reduced peak memory consumption by 1.2\%, with maximum observed savings of 8 MB on hard problems. Importantly, in some cases, MARCO's optimizations slightly increased memory usage (by up to 1.9\%) to achieve runtime speedups, highlighting the deliberate trade-offs between time and space optimizations that characterize effective HPC code.

Table~\ref{tab:memory-results} presents memory usage comparisons across problem difficulties.

\begin{table}[h]
    \centering
    \caption{Average Peak Memory Usage (MB) Across Problem Difficulties}
    \label{tab:memory-results}
    \begin{tabular}{lccc}
    \hline
    \textbf{Model} & \textbf{Easy} & \textbf{Medium} & \textbf{Hard} \\
    \hline
    Claude-3.5 Sonnet & 18.9 & 25.3 & 32.3 \\
    MARCO (ours)      & 18.9 & 25.2 & 28.8 \\
    \textbf{Change} & \textbf{0.0\%} & \textbf{-0.4\%} & \textbf{-10.8\%} \\
    \hline
    \end{tabular}
\end{table}

The memory analysis reveals that MARCO's primary strength lies in runtime optimization rather than memory reduction, with the most significant memory improvements occurring on hard problems where algorithmic optimizations can substantially reduce space complexity.

\subsection{Summary}

Overall, MARCO demonstrated consistent and substantial runtime improvements on hard problems (51.9\% average improvement), competitive performance on medium problems (15.6\% improvement), and modest gains on easy problems (21.2\% improvement). While memory optimization showed less dramatic improvements, MARCO maintained comparable or better memory efficiency across all problem categories. These findings establish MARCO as a promising framework for scalable, architecture-aware code optimization in HPC workloads, with particular effectiveness on complex algorithmic challenges. The results warrant further exploration into cross-model comparisons, reinforcement-based optimization loops, and domain-specific extensions to fully realize the potential of multi-agent code optimization approaches.
\section{Multi-Agent System for Optimizing LLM Code (MARCO)}
\label{marco}

\subsection{Motivation}

The rapid growth of computational demands across industries has driven unprecedented interest in high-performance computing (HPC) applications. With the emergence of large language models (LLMs), there has been a parallel surge in attempts to leverage LLMs for HPC-related code generation and optimization. However, our findings reveal that state-of-the-art LLMs exhibit notable inefficiencies when generating HPC code, motivating the design of a multi-agent system (MAS).

Three key limitations drive our approach. First, a significant constraint arises from the pretraining knowledge cutoff of current LLMs, which restricts their access to the latest advances in HPC architectures, optimization strategies, and software libraries. To address this gap, we integrate a web-search component into our agent system, enabling dynamic retrieval of up-to-date optimization techniques from scholarly resources such as IEEE, ACM, arXiv, and ResearchGate. This allows MARCO to explore and incorporate emerging methods, language updates, and cutting-edge toolchains, surpassing the static knowledge of the underlying LLM.

Second, cost-efficiency presents a major concern. Prior work on fine-tuning monolithic LLMs for HPC code generation demands extensive GPU resources and substantial training overhead~\cite{ding2023hpc}. MARCO leverages pretrained LLMs without requiring fine-tuning, instead enriching them with system-level context and externally sourced insights, making the solution both scalable and accessible.

Third, accessibility barriers limit HPC adoption. Manual optimization, while often effective, requires deep expert knowledge and is labor-intensive, raising the barrier to entry for scientific software developers. By combining up-to-date research knowledge with automated code transformation and iterative feedback, MARCO aims to significantly reduce the need for manual optimization and deliver diverse, performant solutions for large-scale HPC projects, ultimately lowering the barrier to innovation in high-performance software engineering.

\subsection{MARCO Framework}

\begin{figure}[h]
    \centering
    \includegraphics[width=0.5\textwidth]{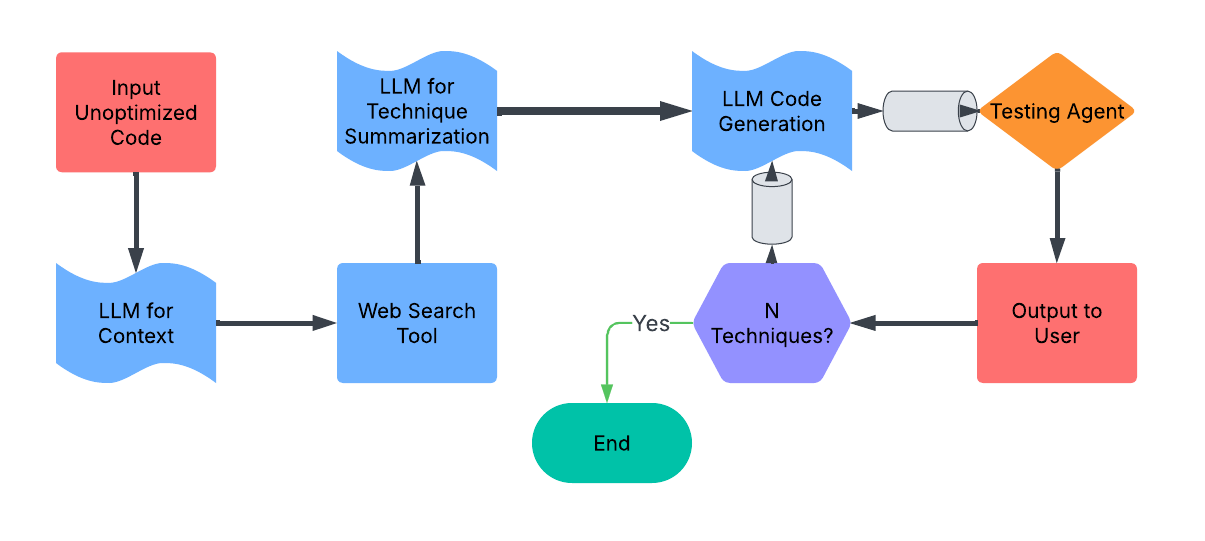}
    \caption{\textbf{MARCO Pipeline Overview.} The iterative optimization process showing the interaction between the code optimizer agent, web search engine, and performance evaluator agent. The system begins with input code, searches for relevant optimization techniques, applies improvements, evaluates performance, and iterates until optimal results are achieved.}
    \label{fig:marco_pipeline}
\end{figure}

The MARCO framework operates through three core components working in a coordinated iterative loop:

\begin{itemize}
    \item \textbf{Code Optimizer Agent:} This agent interprets the input code snippet, generates structured queries for the web-search engine, and applies retrieved optimization strategies to improve memory efficiency, execution performance, and scalability. It performs multiple refinement iterations, progressively integrating feedback from the Performance Evaluator Agent. This iterative design aligns with foundational principles in multi-agent systems~\cite{wooldridge1995intelligent} and mimics multi-step reasoning patterns seen in modern agentic workflows~\cite{huang2023agentcoder}.

    \item \textbf{Web-Search Engine:} We integrate the Tavily search tool~\cite{tavily2024} to retrieve relevant optimization techniques from scholarly sources, including acm.org, ieee.org, arxiv.org, and researchgate.net. The search engine is configured with adjustable parameters including \texttt{search\_depth} (which controls the thoroughness of each search), \texttt{include\_answer} (which determines whether to include direct answers or just source links), and the number of sources retrieved per query. This ensures that the optimizer agent has access to diverse and current references for each optimization iteration.

    \begin{figure}[h]
        \centering
        \includegraphics[width=0.3\textwidth]{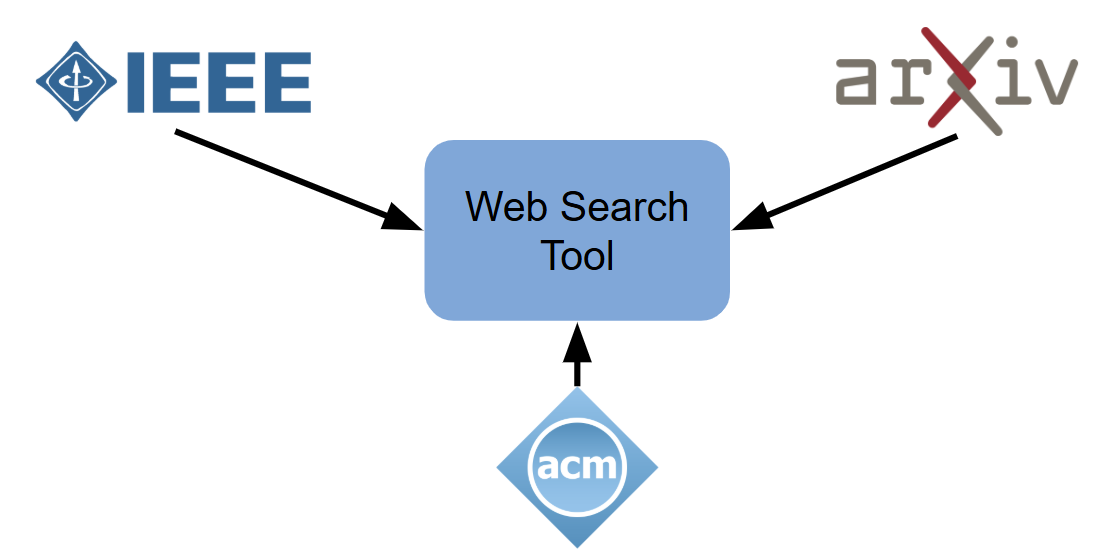}
        \caption{\textbf{Web-Search Integration Architecture.} Schematic demonstrating how external scholarly sources (IEEE, ACM, arXiv, ResearchGate) feed current optimization strategies and techniques into MARCO's iterative optimization loop, bridging the knowledge gap between LLM training cutoffs and current HPC research.}
        \label{fig:web_search}
    \end{figure}

    \item \textbf{Performance Evaluator Agent:} This agent benchmarks the generated code across multiple metrics, including execution time, memory usage, algorithmic complexity, and computational cost. It provides structured feedback to the optimizer agent, enabling the system to iteratively refine solutions based on quantitative performance data. Performance results are presented to users in a tabular interface, summarizing performance trends across iterations, correctness evaluations, and complexity assessments.
\end{itemize}

The three components operate in a closed feedback loop: the optimizer generates improved code based on web-sourced techniques, the evaluator measures performance improvements, and this feedback guides subsequent optimization iterations until convergence or user-defined stopping criteria are met.

\subsection{Evaluation of MARCO}

We evaluated MARCO-optimized code against baseline LLM-generated solutions using the LeetCode 75 dataset~\cite{leetcode75} and an additional set of 10 randomly selected hard problems. Critically, we compared the performance of MARCO with and without the web-search component, benchmarking results against Claude-3.5 Sonnet and GPT-4o to isolate the contribution of dynamic knowledge retrieval.

Across all evaluated problems, MARCO with the Tavily web-search integration consistently outperformed both Claude-3.5 Sonnet and MARCO without web-search enhancement. This demonstrates the substantial value of incorporating real-time, externally sourced optimization strategies, especially in the rapidly evolving HPC landscape where new architectures, libraries, and techniques frequently emerge. The web-search component enables MARCO to access optimization techniques published after the base LLM's training cutoff, providing a significant competitive advantage.

The iterative refinement process contributed significantly to performance gains. On harder LeetCode problems, subsequent iterations generally achieved better results than initial attempts, with the final iteration showing an average improvement of 30.9\% over the base MARCO system without web search. This improvement aligns with our hypothesis that complex optimization problems benefit from multiple rounds of technique application and performance-based refinement.

\begin{figure}[h]
    \centering
    \includegraphics[width=0.3\textwidth]{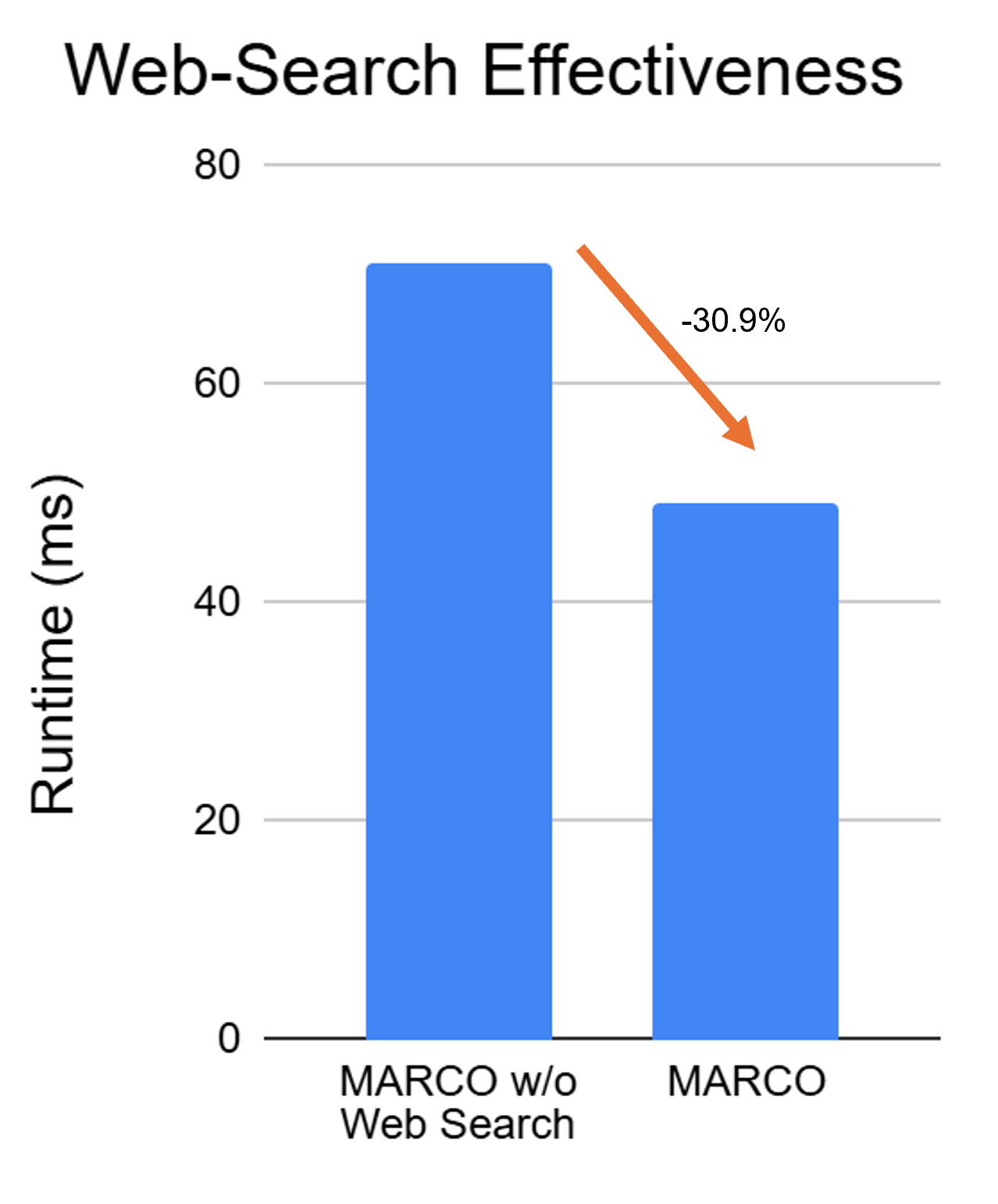}
    \caption{\textbf{Web-Search Component Performance Impact.} Runtime comparison showing performance differences between MARCO with and without the Tavily web-search tool on the LeetCode 75 problem set. The results demonstrate consistent improvements when current optimization techniques are dynamically integrated into the optimization process.}
    \label{fig:websearch_results}
\end{figure}

Figure~\ref{fig:websearch_results} illustrates the runtime performance differences between MARCO with and without the Tavily web-search tool across the evaluated problem set, clearly demonstrating the value of dynamic knowledge integration in automated code optimization.

\section{Discussion}
\label{discussion}

\subsection{Key Findings and Analysis}

Our experiments demonstrate that MARCO substantially enhances execution speed, memory efficiency, and algorithmic optimization compared to baseline LLM-generated code. The most significant finding is that performance improvements scale with problem complexity: while MARCO shows modest gains on simple problems (21.2\% improvement on easy problems), it achieves substantial improvements on complex tasks (51.9\% improvement on hard problems). This scaling behavior aligns with our hypothesis that multi-agent optimization provides the greatest value for computationally demanding scenarios where sophisticated optimization techniques offer substantial benefits.

The integration of the web-search component consistently yields superior optimizations across all problem categories, enabling the system to overcome the inherent knowledge limitations of static pretrained models. By dynamically retrieving cutting-edge techniques from recent research publications, MARCO bridges the critical gap between model training cutoffs and the rapidly evolving landscape of HPC optimization. Our results show a 30.9\% improvement when web search is enabled, demonstrating the substantial value of real-time knowledge integration in automated code optimization.

The iterative refinement process proves particularly effective for complex algorithmic problems, where multiple optimization strategies can be layered and refined based on performance feedback. This multi-step approach mirrors human expert behavior, where initial solutions are progressively improved through systematic analysis and technique application.

\subsection{System Limitations and Trade-offs}

Despite its advantages, MARCO introduces several important trade-offs that must be considered for practical deployment. First, the multi-agent architecture incurs computational overhead through iterative communication between agents and the transmission of intermediate results. This results in slower code generation compared to traditional single-pass LLM workflows, with generation time increasing approximately linearly with the number of optimization iterations.

Second, while the web-search component enables access to recent research, the system's performance remains constrained by the underlying LLM's ability to correctly interpret and implement advanced HPC techniques retrieved from scholarly sources. Complex optimization strategies may be misapplied or inadequately integrated, particularly when they require deep domain expertise or specialized implementation knowledge.

Third, MARCO's current implementation faces several technical constraints. The Performance Evaluator Agent lacks native support for key HPC libraries, including Numba, Torch, and CUDA, significantly limiting optimization potential for GPU-accelerated workloads. Additionally, the system is currently restricted to Python, which, while offering a rich ecosystem of optimization tools, excludes important HPC languages such as C++, Fortran, and specialized domain-specific languages.

Finally, cost considerations present practical limitations. While MARCO's approach avoids expensive fine-tuning, the iterative nature of the optimization process increases API usage costs. Our analysis shows higher per-query costs for more sophisticated models like GPT-4o, despite occasionally inferior performance compared to Claude 3.5 Sonnet, suggesting the need for careful model selection and cost-performance optimization.

\subsection{Unexpected Findings and Implications}

Our evaluation revealed several counterintuitive results that provide important insights for multi-agent system design. Most notably, standalone LLMs occasionally outperformed MARCO on simple problems, particularly those in the "easy" category of the LeetCode 75 dataset. This occurs because simple problems inherently require minimal optimization, and MARCO's multi-agent orchestration introduces unnecessary computational overhead for tasks that are already well-solved by basic implementations.

This finding has important implications for system design: effective automated optimization systems must include problem complexity assessment to determine when sophisticated optimization is warranted. Simple problems may be better served by direct LLM generation, while complex problems benefit from MARCO's iterative refinement approach.

The scaling relationship between problem complexity and optimization benefit suggests that MARCO's architecture is well-suited for its intended domain of high-performance computing, where problems are typically computationally demanding and benefit from sophisticated optimization techniques. This validates our design hypothesis that multi-agent systems provide the greatest value for complex, domain-specific tasks rather than general-purpose code generation.

\subsection{Broader Implications and Future Directions}

These findings have significant implications for the development of AI-assisted programming tools in specialized domains. The success of dynamic knowledge integration through web search suggests that hybrid approaches combining pretrained models with real-time information retrieval may be more effective than purely static or purely fine-tuned approaches for rapidly evolving technical domains.

Future work should address the identified limitations through several key directions: expanding language support beyond Python to include C++ and Fortran, integrating support for GPU acceleration libraries, developing more sophisticated problem complexity assessment mechanisms, and optimizing the cost-performance trade-offs in multi-agent architectures.

Additionally, evaluation on real-world HPC applications beyond algorithmic benchmarks would provide valuable insights into MARCO's practical effectiveness for scientific computing workloads. The development of domain-specific evaluation metrics for HPC code quality, including parallelization efficiency and hardware utilization, represents another important research direction.

The demonstrated effectiveness of MARCO's approach suggests promising applications to other specialized programming domains where current LLMs show limitations, including systems programming, embedded software development, and domain-specific scientific computing applications.
\section{Conclusions and Future Work}
\label{conclusions}

\subsection{Summary of Contributions}

In this work, we addressed the significant limitations of current LLMs in generating optimized HPC code by introducing MARCO, a novel multi-agent system designed to iteratively refine code for improved performance, memory efficiency, and scalability. Our comprehensive evaluation demonstrates that MARCO effectively leverages external knowledge sources and iterative optimization loops to substantially outperform standalone LLMs, achieving a 14.6\% average runtime reduction on the LeetCode 75 problem set and a remarkable 51.9\% improvement on complex problems compared to Claude 3.5 Sonnet.

The integration of real-time web search capabilities proves particularly valuable, yielding a 30.9\% performance improvement over the base MARCO system and enabling access to cutting-edge optimization techniques beyond the knowledge cutoff of pretrained models. These results establish MARCO as an effective approach to automated HPC code optimization that bridges the gap between static LLM capabilities and the rapidly evolving landscape of high-performance computing.

Our key contributions include: (1) a specialized multi-agent architecture that separates code generation, evaluation, and testing responsibilities; (2) an adaptive feedback mechanism that enables iterative code refinement; (3) real-time knowledge integration through web search of scholarly sources; and (4) a cost-effective implementation that avoids expensive fine-tuning while delivering superior performance.

\subsection{Future Research Directions}

Building upon these promising results, several important directions emerge for future development and research. We organize these opportunities into three primary categories: system enhancements, evaluation expansion, and broader applications.

\subsubsection{System Enhancements}

The most immediate improvements focus on enhancing inter-agent communication and expanding system capabilities. Currently, the Performance Evaluator Agent provides feedback solely through final performance metrics. Implementing richer bidirectional communication—including intermediate diagnostic information, specific bottleneck identification, and targeted optimization suggestions—could significantly improve the refinement process efficiency and effectiveness.

Technical infrastructure improvements represent another crucial development area. Integrating Docker-based execution environments would support a wider range of external libraries (including Numba, Torch, and CUDA) and enable more robust, cross-platform testing capabilities. Additionally, extending language support beyond Python to include C++, Fortran, and domain-specific languages would significantly broaden MARCO's applicability to real-world HPC applications.

The optimization workflow itself offers substantial improvement potential. Rather than generating complete code variants for each iteration, MARCO could maintain and incrementally edit persistent code artifacts, enabling finer-grained improvements and reducing redundant computations. Incorporating reinforcement learning techniques, as demonstrated in systems like AlphaCode~\cite{li2022competition}, could enable MARCO to learn from iterative feedback and improve optimization strategies over time~\cite{sun2023reinforcement,sun2024llm}.

\subsubsection{Evaluation and Validation}

Future evaluation efforts should expand beyond algorithmic benchmarks to include real-world HPC applications across diverse scientific domains. Developing domain-specific evaluation metrics that capture HPC-relevant qualities—such as parallelization efficiency, hardware utilization, and scalability characteristics—would provide more comprehensive assessment of optimization effectiveness.

Long-term studies examining MARCO's performance consistency across different hardware architectures and workload types would validate the generalizability of our findings. Additionally, comparative studies with other automated optimization approaches, including compiler optimizations and specialized fine-tuned models, would better position MARCO within the broader optimization landscape.

\subsubsection{Broader Applications and Impact}

The success of MARCO's multi-agent approach suggests promising applications to other specialized programming domains where current LLMs show limitations, including systems programming, embedded software development, and domain-specific scientific computing. The demonstrated effectiveness of dynamic knowledge integration through web search may prove valuable for rapidly evolving technical domains beyond HPC.

Furthermore, integrating persistent communication history across optimization sessions could reduce token usage costs and enable more sophisticated multi-session optimization pipelines, making the approach more practical for large-scale deployment in research and industrial settings.

\subsection{Concluding Remarks}

MARCO represents a significant advancement in automated HPC code optimization, demonstrating that multi-agent orchestration combined with real-time knowledge retrieval can substantially improve upon traditional single-model approaches. By effectively bridging the gap between static pretrained LLMs and the rapidly evolving landscape of high-performance computing, MARCO establishes a foundation for scalable, cost-efficient, and accessible optimization pipelines.

The demonstrated performance improvements, particularly on complex problems where sophisticated optimization provides the greatest benefit, suggest that MARCO's approach addresses a genuine need in the HPC community. As computational demands continue to grow across scientific and industrial domains, automated optimization tools like MARCO have the potential to significantly lower barriers to high-performance programming and accelerate innovation in computationally intensive applications.

The path forward involves continued development of the technical capabilities outlined above, broader evaluation across real-world applications, and exploration of the approach's applicability to other specialized programming domains. Through these efforts, MARCO's multi-agent optimization paradigm may contribute to making high-performance computing more accessible and effective for the broader scientific and engineering communities.

\printbibliography

\end{document}